%% file: main.tex
\documentclass[submission]{eptcs}

\usepackage{cite}

\usepackage{graphicx}
\usepackage{bussproofs}
\usepackage{dsfont}
\usepackage{times}
\usepackage{mathtools}              
\mathtoolsset{showonlyrefs}  
\usepackage{url}
\usepackage{verbatim}
\usepackage{xargs}
\usepackage{xspace}
\usepackage{MnSymbol}
\usepackage{enumerate}
\usepackage{yfonts}
\usepackage{paralist}
\usepackage{upgreek}
\usepackage{multirow}

\usepackage{tikz}
\usetikzlibrary{arrows,shadows,matrix,shapes,positioning,chains,calc}

\usepackage{subfigure}
\usepackage{example/figures}

\input{macros}

\newcommand{\titlerunning}{Communicating machines as a dynamic binding mechanism of services}
\newcommand{\authorrunning}{Vissani, Lopez Pombo, Tuosto}

\begin{document}

\title{
  Communicating machines as a dynamic binding mechanism of services\thanks{This work has been supported by the
    European Union Seventh Framework Programme under grant agreement
    no. 295261 (MEALS)}
}

\author{Ignacio Vissani\institute{Department of computing \\ School of Science \\ Universidad de Buenos Aires}\email{ivissani@dc.uba.ar} \and Carlos Gustavo Lopez Pombo\institute{Department of computing \\ School of Science \\ Universidad de Buenos Aires \and Consejo Nacional de Investigaciones \\ Cient\'ificas y Tecnol\'ogicas}\email{clpombo@dc.uba.ar} \and Emilio Tuosto\institute{Department of Computer Science \\ University of Leicester}\email{emilio@leicester.ac.uk}}


\maketitle

\begin{abstract}
Distributed software is becoming more and more dynamic to
support applications able to respond and adapt to the changes of their
execution environment. For instance, \emph{service-oriented computing} (SOC) envisages
applications as services running over globally available computational
resources where discovery and binding between them is transparently performed by a middleware.
\emph{Asynchronous Relational Networks} (ARNs) is a well-known formal orchestration model, based on
hypergraphs, for the description of service-oriented software artefacts.
Choreography and orchestration are the two main design principles for
the development of distributed software. 
In this work, we propose \emph{Communicating Relational Networks} (CRNs), which is a variant of ARNs, but relies on choreographies for the characterisation of the communicational aspects of a software artefact, and for making their automated analysis more efficient.
\end{abstract}

\input{introduction}
\input{preliminaries}
\input{def_example}
\input{crns}
\input{conc}

\bibliographystyle{eptcs}
\bibliography{bibdatabase}


\end{document}

%% file: macros.tex
\newcommand{\Q}{\mathbf{\mathsf{Q}}}
\newcommand{\C}{\mathbf{\mathsf{C}}}
\newcommand{\msgs}{\mbox{\textcolor{blue}{$\mathsf{Msg}$}}}
\newcommand{\amsg}[1][m]{\textcolor{blue}{\mathsf{#1}}}
\newcommand{\aport}[1][{}]{\textcolor{blue}{\pi^{#1}}}

\newtheorem{definition}{Definition}

\newcommand{\Sit}{\textcolor{blue}{\ensuremath{\mathit{S}}}}
\newcommand{\vectorize}[1]{\ensuremath{\overset{\rightarrow}{#1}}}
\newcommand{\sysconf}[2]{\ensuremath{(\vectorize{#1}, \vectorize{#2})}}

\newcommand{\cfsmalphabet}{\msgs}
\newcommand{\cfsmfinitewords}{\ensuremath{\cfsmalphabet^\star}}

\newcommand{\machine}[1]{\ensuremath{\mathsf{#1}}\xspace}

\newcommand{\FS}{\ensuremath{\mathcal{F}}\xspace}

\newcommand{\PSet}{\textcolor{orange}{\mathsf{P}}}

\newcommand{\ptp}[1]{\mathsf{#1}}
\newcommand{\p}{\ptp p}
\newcommand{\q}{\ptp q}

\newcommand{\chan}[1]{\ptp{#1}}
\newcommand{\PSEND}[2]{\chan{#1} \oact \msgsort{#2}}
\newcommand{\PRECEIVE}[2]{\chan{#1} \iact \msgsort{#2}}
\newcommand{\oact}{\ptp !}
\newcommand{\iact}{\ptp ?}
\newcommand{\msgsort}[1]{\mathsf{#1}}

\newcommand{\conf}[1]{\langle #1 \rangle}

\newcommand{\defref}[2][{}]{Definition{#1}~\ref{#2}}
\newcommand{\secref}[2][{}]{Section{#1}~\ref{#2}}

\newcommand{\LSet}{\mathsf{L}}

%% file: introduction.tex
\section{Introduction and motivation}
\label{introduction}

Distributed software is becoming more and more dynamic to
support applications able to respond and adapt to the changes of their
execution environment.
For instance, \emph{service-oriented computing} (SOC) envisages
applications as services running over globally available computational
resources; at run-time, services search for other services to bind to
and use.
Software architects and programmers have no control as to the nature
of the components that an application can bind to due to the fact that
the discovery and binding are transparently performed by a middleware.

Choreography and orchestration are the two main design principles for
the development of distributed software (see e.g.,~\cite{pel03}). Coordination is attained in
the latter case by an \emph{orchestrator}, specifying (and possibly
executing) the distributed work-flow. Choreography features the notion
of \emph{global view}, that is a holistic specification describing
distributed interactions amenable of being ``projected'' onto the
constituent pieces of software.
In an orchestrated model, the distributed computational components
coordinate with each other by interacting with a special component,
\emph{the orchestrator}, which at run time decides how the work-flow
has to evolve.
For example the orchestrator of a service offering the booking of a
flight and a hotel may trigger a service for hotel and one for flight
booking in parallel, wait for the answers of both sites, and then
continue the execution.
In a choreographed model, the distributed components autonomously
execute and interact with each other on the basis of a local control
flow expected to comply with their role as specified in the ``global
viewpoint''.
For example, the choreography of hotel-flight booking example above
could specify that the flight service interacts with the hotel service
which in turns communicates the results to the buyer.

We use Asynchronous relational networks (ARNs)~\cite{tutu:calco2013}
as the basis of our approach.
In ARNs, systems are formally modelled as hypergraphs obtained by
connecting hyperarcs which represent units of computation and
communication.
More precisely, hyperarcs are interpreted as either processes
(\emph{services} or unit of computation) or as communication
channels (unit of communication).
The nodes can only be adjacent to:
\begin{inparaenum}[1.]
\item one process hyperarc and one communication hyperarc, meaning that the computation formalised by the process hyperarc is bound through the communication channel formalised by the communication hyperarc,
\item one process hyperarc, meaning that it is a \emph{provides-point} through which the computation formalised by the process hyperarc can be bound to an activity that requires that particular service, or
\item one communication hyperarc, meaning that it is a \emph{requires-point} to which a given service can be bound using one of its provides-points.
\end{inparaenum}
The rationale behind this separation is that a provides-point yields
the interface through which a service exports its functionality while
a requires-point is the interface through which an activity expects
certain service to provide a functionality.
Composition of services can then be understood as fusing a provides-point with a requires-point in a way that the service exported by the former satisfies the expectations of the latter, usually formalised as contracts in some formal language.

Hyperarcs are labelled with (Muller) automata; in the case
of process hyperarcs, automata formalise the interactions carried out
by that particular service while, in the case of communication
hyperarcs, they represent the orchestrator coordinating the behaviour
of the participants of the communication.
%
%
In fact, the automaton $\Lambda$ associated to a communication
hyperarc coordinates the processes bound to its ports by, at each
time, interacting with one of the processes and deciding, depending
on the state $\Lambda$ is in, what is the next interaction (if any)
to execute.
The global behaviour of the system is then obtained by composing the automata
associated to process and communication hyperarcs.
In the forthcoming sections we will introduce a running example to show how definitions work and concretely discuss the contributions of the present work.

As anticipated, the composition of ARNs yields a semantic definition
of a binding mechanism of services in terms of ``fusion'' of
provides-points and requires-points.
Once coalesced, the nodes become ``internal'', that is they are no
longer part of the interface and cannot be used for further bindings.
In existing works, like~\cite{tutu:calco2013}, the binding is subject to an entailment relation between \emph{linear temporal logic}~\cite{pnueli:tcs-13_1} theories attached to the provides- and requires-points that can be checked by resorting to any decision procedure for LTL (for example,~\cite{kesten:cav93})

Although the orchestration model featured by ARNs is rather expressive
and versatile, we envisage two drawbacks:
\begin{enumerate}
\item the binding mechanism based on LTL-entailment establishes an
  asymmetric relation between requires-point and provides-point as it
  formalises a notion of trace inclusion; also,
\item including explicit orchestrators (the automata labelling the communication hyperarcs), in the composition, together with the computational units (the automata labelling the process hyperarcs)
  increases the size of the resulting automaton making the analysis
  more expensive.
\end{enumerate}

In the present work we propose \emph{Communicating Relational Networks} (CRNs), a variant of ARNs relying on choreographies to overcome those issues, where provides-points are labelled with \emph{Communicating Finite State Machines}~\cite{bz83} declaring the behaviour (from the communication perspective) exported by the service, and communication hyperarcs are labelled with \emph{Global Graphs}~\cite{dy12} declaring the global behaviour of the communication channel.
In this way, our proposal blends the orchestration framework of ARNs with a choreography model based on global graphs and communicating machines. Unlike most of the approaches in the literature (where choreography and orchestration are considered antithetical), we follow a comprehensive approach showing how choreography-based mechanisms could be useful in an orchestration model.

The present work is organised as follows; in \secref{preliminaries}
we provide the formal definitions of most of the concepts used along
this paper.
Such definitions are illustrated with a running example introduced in \secref{sec:defex}.
In \secref{crns} we introduce the main contribution of this paper,
being the definition of CRNs, we show how they are used to rewrite the
running example and we discuss several aspects regarding the
design-time checking to assert internal coherence of services, the
run-time checking ruling the binding mechanism and the cost of
software analysis.
Finally, in \secref{conc} we draw some conclusions and discuss some
further research directions.

%% file: preliminaries.tex
\section{Preliminaries}
\label{preliminaries}
In this section we present the preliminary definitions used throughout
the rest of the present work.
We first summarise communicating machines and global graphs borrowing
definitions from~\cite{lty15} and from~\cite{dy12}.
Finally we introduce some basic definitions in order to present ARNs;
the definitions here are adapted from~\cite{tutu:calco2013}.

\input{def_cfsm}
\input{def_arn}


%% file: def_cfsm.tex
\subsection{Communicating machines and global graphs} 
\label{sub:communicating_machines}
Communicating machines were introduced in~\cite{bz83} to model and
study communication protocols in terms of finite transition systems
capable of exchanging messages through some channels.
We fix a finite set \msgs\ of \emph{messages} ans a finite set $\PSet$ of
participants.
\begin{definition}[\cite{bz83}]
  A \emph{communicating finite state machine} on $\cfsmalphabet$ (CFSMs, for
  short) is a finite transition system 
  $\left(\Q, \C, q_0, \cfsmalphabet, \delta\right)$ where
  \begin{itemize}
  \item $\Q$ is a finite set of states;
  \item $\C = \{ \p\q \in \PSet^2 \st \p \not= \q\}$
    is a set of channels;
  \item $q_0 \in \Q$ is an initial state;
  \item
    $\delta \subseteq \Q \times (\C \times \{!,?\} \times \msgs) \times
    \Q$ is a finite set of \emph{transitions}.
  \end{itemize}
  A \emph{communicating system} is a map $\Sit$ assigning a CFSM
  $\Sit(\p)$ to each $\p \in \PSet$.
  We write $q \in \Sit(\p)$ when $q$ is a state of the machine
  $\Sit(\p)$ and likewise and $t \in \Sit(\p)$ when $t$ is a
  transition of $\Sit(\p)$.
\end{definition}

The execution of a system is defined in terms of transitions between
configurations as follows:
\begin{definition}
  The \emph{configuration} of communicating system $\Sit$ is a pair
  $s = \sysconf{q}{w}$ where
  $\vectorize{q} = \left(q_\p\right)_{\p \in \PSet}$ where
  $q_\p \in \Sit(\p)$ for each $\p \in \PSet$ and
  $\vectorize{w} = \left(w_{\p\q}\right)_{\p\q \in \C}$ with
  $w_{\p\q} \in \cfsmfinitewords$.
  A configuration $s' = \sysconf{q'}{w'}$ is \emph{reachable} from another
  configuration $s = \sysconf{q}{w}$ by the \emph{firing of the
    transition} $t$ (written $s \overset{t}{\rightarrow} s'$) if there
  exists $\amsg \in \cfsmalphabet$ such that either:
	\begin{enumerate}
		\item $t = (q_\p, \p\q!\amsg,  q'_\p) \in \delta_\p$ and 
			\begin{enumerate}
				\item $q'_{\p'} = q_{\p'}$ for all $\p' \not= \p$; and
				\item $w'_{\p\q} = w_{\p\q} \cdot \amsg$ and $w'_{\p'\q'} = w_{\p'\q'}$ for all $\p'\q' \not= \p\q$; or
			\end{enumerate}
		\item $t = (q_\q, \p\q?\amsg,  q'_\q) \in \delta_\q$ and 
			\begin{enumerate}
				\item $q'_{\p'} = q_{\p'}$ for all $\p' \not= \q$; and
				\item $\amsg \cdot w'_{\p\q} = w_{\p\q}$ and $w'_{\p'\q'} = w_{\p'\q'}$ for all $\p'\q' \not= \p\q$
			\end{enumerate}
	\end{enumerate}
\end{definition}

A \emph{global graph} is a finite graph whose nodes are labelled over
the set $\LSet = \{\bigcircle,\bigocirc, \diamondplus, \boxvert \} \
\cup \ \{\ptp s \to \ptp r:\amsg \mid \ptp s,\ptp r \in \PSet \wedge
\amsg \in \cfsmalphabet\}$ according to the following definition.
\begin{definition}
  A \emph{global graph} (over $\PSet$ and $\cfsmalphabet$) is a
  labelled graph $\langle V,A,\Lambda \rangle$ with a set of
  \emph{vertexes} $V$, a set of \emph{edges} $A \subseteq V \times V$,
  and \emph{labelling function} $\Lambda : V \to \LSet$ such that
  $\Lambda^{-1}(\bigcircle)$ is a singleton and, for each $v \in V$
  \begin{enumerate}
  \item if $\Lambda(v)$ is of the form $\ptp s \to \ptp r:\amsg$ then
    $v$ has a unique incoming and unique outgoing edges,
  \item if $\Lambda(v) \in \{\diamondplus, \boxvert\}$ then $v$ has at
    least one incoming edge and one outgoing edge and,
  \item $\Lambda(v) = \bigocirc$ then $v$ has zero outgoing edges.
  \end{enumerate}
\end{definition}
Label $\ptp s \to \ptp r:\amsg$ represents an interaction where machine
$\ptp s$ sends a message $\amsg$ to machine $\ptp r$. A vertex with label $\bigcircle$
reperesents the source of the global graph, $\bigocirc$ represents the
termination of a branch or of a thread, $\boxvert$ indicates forking
or joining threads, and $\diamondplus$ marks vertexes corresponding to
branch or merge points, or to entry points of loops.

In the following we use a projection algorithm that given a global graph
retrieves communicating machines for each of its participants.
Undestranding such algorithm is not necessary for the sake of this paper
and the interested reader
is referred to~\cite{lty15} for its definition.

%% file: def_arn.tex
\subsection{Asynchronous relational networks} \label{sec:arns}
A Muller automaton is a finite state automaton where final states
are replaced by a family of states to define an acceptance condition
on infinite words.
\begin{definition}[Muller automaton]\label{def:muller-automaton}
  A \emph{Muller automaton} over a finite set $A$
  of \emph{actions} is a structure of the form
  $\langle Q, A, \Delta, I, \FS \rangle$ , where
  \begin{enumerate}
  \item $Q$ is a finite set (of \emph{states})
  \item $\Delta \subseteq Q \times A \times Q$
    is a \emph{transition relation} (we write
    $p \xrightarrow{\iota} q$ when $(p, \iota, q) \in \Delta$),
  \item $I \subseteq Q$ is the set of \emph{initial states}, and
  \item $\FS \subseteq 2^Q$ is the set of \emph{final-state sets}.
  \end{enumerate}
  We say that an automaton \emph{accepts} an inifinite trace
  $\omega = q_0~\xrightarrow{\iota_0}~q_1~\xrightarrow{\iota_1}~\ldots$
  if and only if $q_0 \in I$ and there exists $i \geq 0$ and $S \in \FS$ such that for all $s \in S$, the set $\bigcup_{i \leq j \land q_j = s} \{j\}$ is infinite.
\end{definition}

Asynchronous relational networks are hypergraphs connecting
\emph{ports} that can be thought of as communication end-points
through which messages can be sent to or received from other
ports.
\begin{definition}[Port]
  A \emph{port} is a structure $\aport = \conf{\aport[+], \aport[-]}$
  where $\aport[+], \aport[-]$ are disjoint finite sets of messages.
  We say that two ports are disjoint when they are formed by
  componentwise disjoint sets of messages.
  The \emph{actions over $\aport$} are
  $A_{\aport} = \{\pact{m} \st m \in \aport[-]\} \cup \{\dact{m} \st m
  \in \aport[+]\}$.
\end{definition}

The computational agents of ARNs are \emph{processes} formalised as a
set of ports togetherr with a Muller automaton describing the
communication pattern of the agents.
\begin{definition}[Process]
  A \emph{process} $\abr{\gamma, \Lambda}$
  consists of a set $\gamma$
  of pairwise disjoint ports and a Muller automaton $\Lambda$
  over the set of actions
  $A_{\gamma} = \bigcup_{\aport \in \gamma} A_{\aport}$.
\end{definition}

Processes are connected through \emph{connections} whose basic role
is to establish relations among the messages that processes exchange
on the ports of processes and communication hyperedges.
Intuitively, one can thing of the messages used by processes and
communication hyperedges as 'local' messages whose 'global' meaning is
established by connections.
\begin{definition}[Connection]
  Given a set of pairwise disjoint ports $\gamma$, an \emph{attachment
    injection} on $\gamma$ is a pair $\conf{M,\mu}$ where and a finite
  set $M$ of messages and $\mu = \{\mu_{\aport}\}_{\aport \in \gamma}$
  is a family of finite partial injections $\mu_{\aport} \colon M \pto
  \aport[-] \cup \aport[+]$.
  We say that $\abr{M, \mu, \Lambda}$ is a \emph{connection} on
  $\gamma$ iff $\conf{M,\mu}$ is an attachment injection on $\gamma$
  and a Muller automaton $\Lambda$ over $\cbr{\pact{m} \st m
    \in M} \cup \cbr{\dact{m} \st m \in M}$  such that:
  \[
  \mu_{\aport}^{-1}\rbr{\aport[-]} \subseteq \bigcup_{\hat{\aport} \in
  \gamma \setminus \cbr{\aport}} \mu_{\hat{\aport}}^{-1}\rbr{\hat{\aport}^+}
  \qquad \text{and} \qquad
  \mu_{\aport}^{-1}\rbr{\aport[+]} \subseteq \bigcup_{\hat{\aport} \in
  \gamma \setminus \cbr{\aport}} \mu_{\hat{\aport}}^{-1}\rbr{\hat{\aport}^-}.
  \]
 for each $\aport \in \gamma$.
\end{definition}

\begin{definition}[Asynchronous Relational Network~\cite{tutu:calco2013}]
  \label{def:arn}
  Let $M$ be a finite set of messages.
  An \emph{asynchronous relational net} $\alpha$ on $M$ is a structure
  $\abr{X, P, C, \{\aport[]_x\}_{x \in X}, \{\mu_c\}_{c \in C},
    \{\gamma\}_{x \in X}, \{\Lambda_e\}_{e \in P \cup C}}$ consisting of
  \begin{itemize}
  \item a hypergraph $\abr{X, E}$, where $X$ is a (finite) set of \emph{points} and $E = P \cup C$ is a set of \emph{hyperedges} (non-empty subsets of $X$) partitioned into \emph{computation hyperedges} $p \in P$ and \emph{communication hyperedges} $c \in C$ such that no adjacent hyperedges belong to the same partition, and
  \item three labelling functions that assign
    \begin{inlinenum}

    \item a port $\aport[]_{x}$ with messages in $M$ to each point $x \in X$,

    \item a process $\abr{\gamma_{p}, \Lambda_{p}}$ to each hyperedge
      $p \in P$ such that $\gamma_p \subseteq \{\aport[]_x\}_{x \in X}$, and
    \item a connection $\abr{M_{c}, \mu_{c}, \Lambda_{c}}$ to each
      hyperedge $c \in C$.

    \end{inlinenum}
    
  \end{itemize}
An ARN with no provides-point is called \emph{activity} and formalises the notion of a software artefact that can execute, while an ARN that has at least one provides-point is called a \emph{service} and can only execute provided it is bound through one of them to a requires-point of an \emph{activity}.
\end{definition}


%% file: def_example.tex

\section{The running example}\label{sec:defex}
The following running example will help us to present intuitions
behind the definitions, and later, to introduce and motivate our
contributions.
Consider an application providing the service of hotel reservation and
payment processing.
A client activity $\processname{TravelClient}$
asks for hotel options made available by a provider
$\processname{HotelsService}$ returning a list of offers.
If the client accepts any of the offers, then
$\processname{HotelsService}$ calls for a payment processing service
$\processname{PaymentProcessService}$ which will ask the client for
payment details, and notify $\processname{HotelsService}$ whether the
payment was accepted or rejected.
Finally, $\processname{HotelsService}$ notifies the outcome of the
payment process to the client.

\begin{figure}
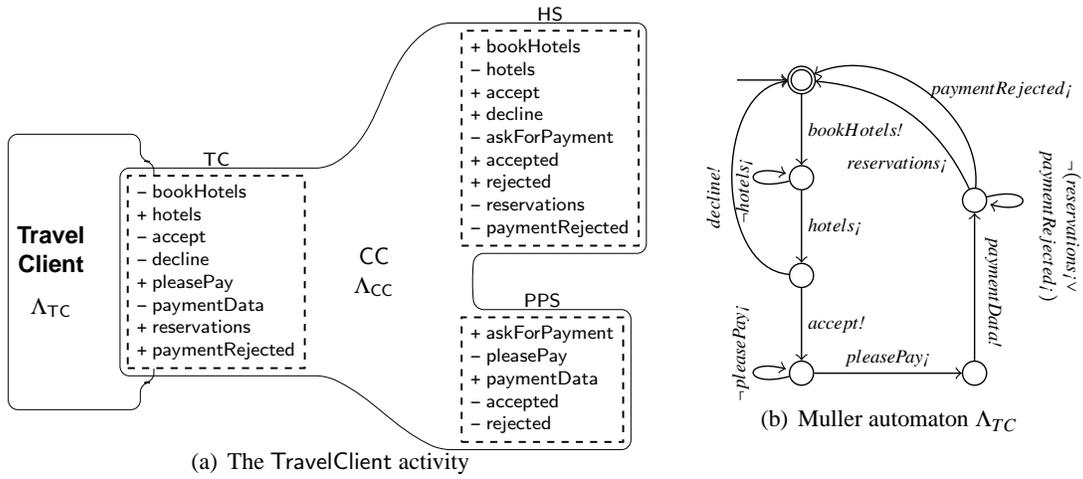
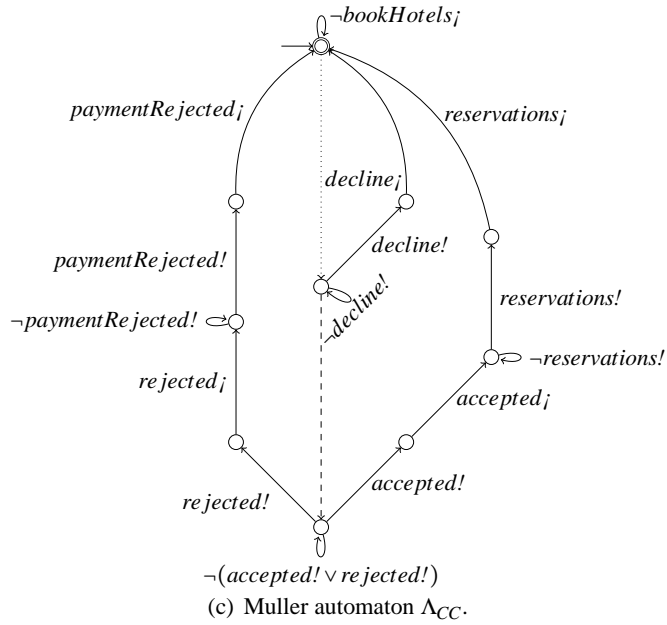

\begin{center}
	\subfigure[The $\processname{TravelClient}$ activity]{
		\scalebox{0.9}{
			\travelClientProcess{\Lambda}
		}
	}
	\subfigure[Muller automaton $\Lambda_{TC}$]{
		\scalebox{1.3}{
			\travelClientAutomaton
		}
	}
	\subfigure[Muller automaton $\Lambda_{CC}$.]{
		\scalebox{0.8}{
			\clientConnectionAbbrAutomaton
		}
	}
	\caption{The $\processname{TravelClient}$ activity together with the Muller automata.}
	\label{figure:travel-client}
\end{center}
\end{figure}

\begin{figure}
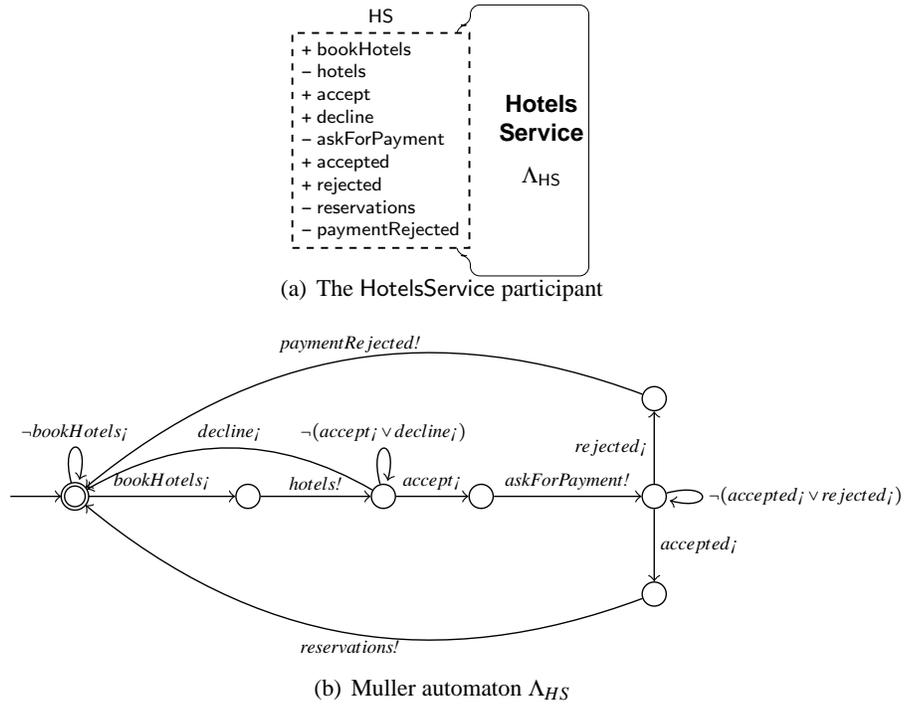

\begin{center}
	\subfigure[The $\processname{HotelsService}$ participant]{
		\scalebox{0.9}{
			\hotelsService 
		}
	}
	\subfigure[Muller automaton $\Lambda_{HS}$]{
		\scalebox{1.3}{
			\hotelsServiceAutomaton
		}
	}
	\caption{The $\processname{HotelsService}$ participant together with the machine \machine{Hs}}
	\label{figure:hotels-service}
\end{center}
\end{figure}

\begin{figure}
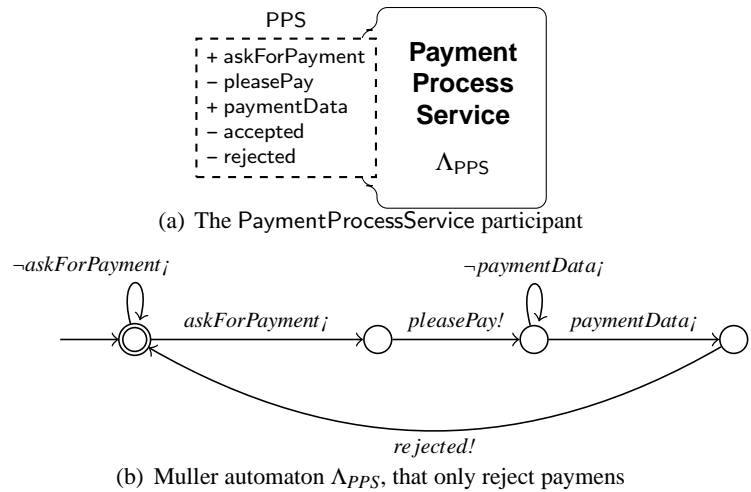

\begin{center}
	\subfigure[The $\processname{PaymentProcessService}$ participant]{
          \begin{minipage}[c]{.5\linewidth}\center
            \scalebox{1}{ \paymentProcessService }
          \end{minipage}
	}
	\subfigure[Muller automaton $\Lambda_{PPS}$, that only reject paymens]{
		\scalebox{1.5}{
			\paymentProcessServiceAutomaton
		}
	}
	\caption{The $\processname{PaymentProcessService}$ participant.}
	\label{figure:payment-service}
\end{center}
\end{figure}

Figures~\ref{figure:travel-client}, \ref{figure:hotels-service},
and~\ref{figure:payment-service} show the ARNs (including the
automata), for the $\processname{TravelClient}$,
$\processname{HotelsService}$, and
$\processname{PaymentProcessService}$ respectively.
The ARN in Fig.~\ref{figure:travel-client}(a) represents an
activity composed with a communication channel.
More precisely, $\processname{TravelClient}$ (in the solid box on the
left) represents a process hyperedge whose Muller automaton is
$\Lambda_{TC}$ (depicted in Fig.~\ref{figure:travel-client}(b)).
The solid ``y-shaped'' contour embracing the three dashed boxes
represents a communication hyperedge used to specify the two
requires-points (i.e., $\portname{HS}$ and $\portname{PPS}$) of the
component necessary to fulfill its goals.
Note that such ARN does not provide itself any service to other
components and that the dashed box lists the outgoing and incoming
messages expected (respectively denoted by names prefixed by '+' and
'-' signs).

It is worth remarking that communication hyperarcs in ARNs yield
the coordination mechanism among a number of services.
In fact, a communication hyperarc enables the interaction among the
services that bind to its requires-points such as
$\processname{TravelClient}$, $\processname{HotelsService}$, and
$\processname{PaymentProcessService}$ in our example.
The coordination is specified through a Muller automaton associated
with the communication hyperarc that acts as the orchestrator of the
services.
In our running example, the communication hyperarc of
Fig.~\ref{figure:travel-client} is labeled with the automaton
$\Lambda_{CC}$ of Fig.~\ref{figure:travel-client}(c) where, for
readability and conciseness, the dotted and dashed edges stand for the
paths
\begin{eqnarray*}
  & \mexplained
  \\ \text{and} \\
  & \nexplained
\end{eqnarray*}
respectively.
As we will see, such automaton corresponds to a global choreography when
replacing the binding mechanism of ARNs with choreography-based mechanisms.
The transitions of the automata are labelled with input/output actions; according
to the usual ARNs notation, a label $\pact{m}$ stands for the ouput of message $m$
while label $\dact{m}$ stands for the input of message $m$.

Figures~\ref{figure:hotels-service}~and~\ref{figure:payment-service}
represent two services with their automata (resp. $\Lambda_{HS}$ and
$\Lambda_{PPS}$) and their provides-point (resp. $\portname{HS}$ and
$\portname{PPS}$) not bound to any communication channel yet.

The composition of ARNs yields a semantic definition of a
binding mechanism of services in terms of ``fusion'' of
provides-points and requires-points.
More precisely, the binding is subject to an entailment relation
between \emph{linear temporal logic}~\cite{pnueli:tcs-13_1} theories
attached to the provides- and requires-points as illustrated in the
following section.


%% file: crns.tex
\section{Communicating Relational Networks}
\label{crns}
As we mentioned before, even when the orchestration model featured by ARNs is rather expressive
and versatile, we envisage two drawbacks which now can be presented in more detail.

\subsection{On the binding mechanism}
If we consider the binding mechanism based on LTL entailment presented in previous works, the relation between requires-point and provides-point is established in an asymmetric way whose semantics is read as trace inclusion. This asymmetry leads to undesired situations. 
For instance, if we return to our running example, a contract stating that
the outcome of an execution is either $\mathit{accept}$ or $\mathit{reject}$ of a payment could be specified by assigning the LTL formula 
\[
\Diamond ((-\mathit{accept} \lor -\mathit{reject}) \land \neg(-\mathit{accept} \land -\mathit{reject}))
\]
to the requires-point ${\sf PPS}$ of Fig.~\ref{figure:travel-client}(a).
Likewise, one could specify a contract for the provides-point ${\sf
  PPS}$ of the ARN in Fig.~\ref{figure:payment-service}(b) stating
that payments are always rejected by including the formula\footnote{In these examples we
  use two propositions, ${\it accept}$ and ${\it reject}$, forcing us
  to include in the specification their complementary behaviour, but
  making the formulae easier to read.}
\[
\Diamond (-{\it reject} \land \neg-{\it accept})
\]
It is easy to show that
\begin{eqnarray*}
\Diamond (-{\it reject} \land \neg-{\it accept}) & \vdash^{\it LTL} &
\Diamond ((-{\it accept} \lor -{\it reject}) \land \neg(-{\it accept}
\land -{\it reject}))
\end{eqnarray*}
by resorting to any decision procedure for LTL
(see for instance,~\cite{kesten:cav93}). The intuition is that every state
satisfying $-{\it reject} \land \neg-{\it accept}$ also satisfies
$(-{\it accept} \lor -{\it reject}) \land \neg(-{\it accept} \land
-{\it reject})$ so if the former eventually happens, then also the
latter.

The reader should note that this scenario leads us to accept a service provider that, even when it can appropriately ensure a subset of the expected outcomes, cannot guaranty that all possible outcomes will eventually be produced.

\medskip

\emph{Communicating Relational Networks} are defined exactly as ARNs but with the definition of \emph{Connection} based on global graphs where, given a set of ports, the messages are related to the messages in the ports, and the participants are identified by the ports themselves.

\begin{definition}[Connection]
  We say that $\abr{M, \mu, \Gamma}$ is a \emph{connection} on
  $\gamma$ iff $\conf{M,\mu}$ is an attachment injection on $\gamma$
  and $\Gamma$ is a global graph where the set of participants is
  $\{\ptp p_\pi\}_{\pi \in \gamma}$ exchanging messages in $M$ such
  that:
  \[
  \mu_{\aport}^{-1}\rbr{\aport[-]} \subseteq \bigcup_{\hat{\aport} \in
  \gamma \setminus \cbr{\aport}} \mu_{\hat{\aport}}^{-1}\rbr{\hat{\aport}^+}
  \qquad \text{and} \qquad
  \mu_{\aport}^{-1}\rbr{\aport[+]} \subseteq \bigcup_{\hat{\aport} \in
  \gamma \setminus \cbr{\aport}} \mu_{\hat{\aport}}^{-1}\rbr{\hat{\aport}^-}.
  \]
 for each $\aport \in \gamma$.
\end{definition}

\begin{definition}[Communicating relational network]
\label{def:crn}
  A \emph{communicating relational net} $\alpha$ is a structure $\abr{X, P, C, \gamma, M, \mu, \Lambda}$ consisting of:
  \begin{itemize}
    
  \item a hypergraph $\abr{X, E}$, where $X$ is a (finite) set of \emph{points} and $E = P \cup C$ is a set of \emph{hyperedges} (non-empty subsets of $X$) partitioned into \emph{computation hyperedges} $p \in P$ and \emph{communication hyperedges} $c \in C$ such that no adjacent hyperedges belong to the same partition, and

  \item three labelling functions that assign
    \begin{inlinenum}

    \item a port $M_{x}$ to each point $x \in X$,

    \item a process $\abr{\gamma_{p}, \Lambda_{p}}$ to each hyperedge $p \in P$, and

    \item a connection $\abr{M_{c}, \mu_{c}, \Lambda_{c}}$ to each hyperedge $c \in C$.

    \end{inlinenum}
    
  \end{itemize}
\end{definition}

\begin{figure}
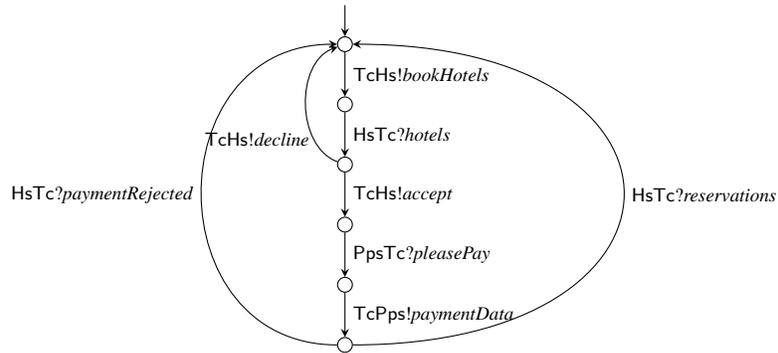
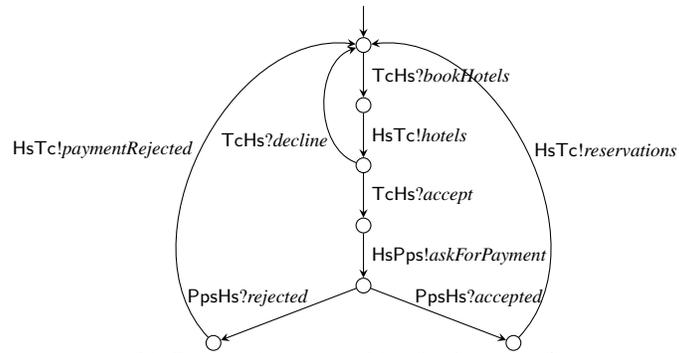
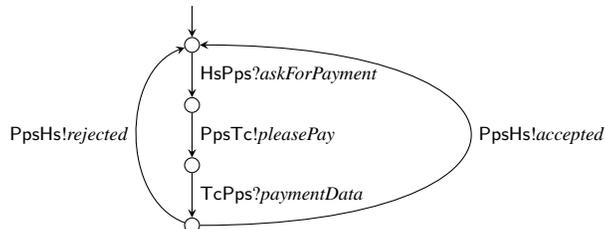

\begin{center}
	\subfigure[Communicating machine for the port $\portname{TC}$]{
	\scalebox{1}{
		\travelClientMachine
	}
	}
\\
	\subfigure[Communicating machine for the port $\portname{HS}$]{
	\scalebox{1}{
		\hotelsServiceMachine
	}
	}
	\subfigure[Communicating machine for the port $\portname{PPS}$]{
	\scalebox{1}{
		\paymentProcessServiceRightMachine
	}
	}
	\caption{Communicating machines labelling the ports  $\portname{TC}$, $\portname{HS}$ and $\portname{PPS}$.}
	\label{figure:communicating-machines}
\end{center}
\end{figure}

\begin{figure}
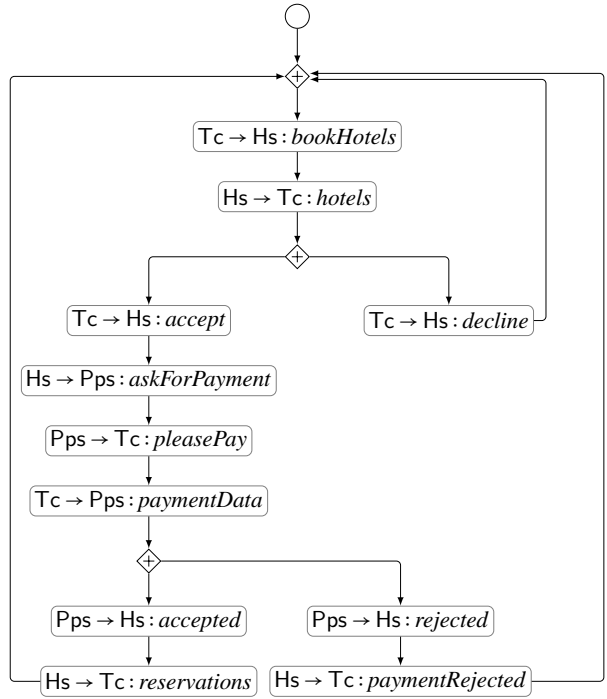

\begin{center}
	\scalebox{.8}{
          \clientConnectionGlobalGraph
	}
	\caption{Global graph of the running example}
	\label{figure:global-graph}
\end{center}
\end{figure}

Figures~\ref{figure:communicating-machines}~and~\ref{figure:global-graph} show the communicating machines and global graphs that can be used to redefine the same services of the running example presented in Sec.~\ref{preliminaries}, but as CRNs.

The machine in Fig.~\ref{figure:communicating-machines}(a) specifies that upon
reception of a \textit{bookHotel} message from the client,
$\processname{HotelsService}$ sends back a list of \textit{hotels}; if
the client accepts then computation continues, otherwise the
$\processname{HotelsService}$ returns to its initial state, etc..
Also, Figs.~\ref{figure:communicating-machines}(b)~and~(c) depict
the communicating machines associated to the provides-points of
services $\processname{HotelsService}$ and
$\processname{PaymentProcessService}$, respectively.
From the point of view of the requires-points, the expected behaviour
of the participants of a communication is declared by means of a
choreography associated to communication hyperarcs.
We illustrate such graphs by discussing the choreography in
Fig.~\ref{figure:global-graph} (corresponding to the automaton in
Fig.~\ref{figure:travel-client}(c)).
The graph dictates that first client and
$\processname{HotelsService}$ interact to make the request and
receive a list of available hotels, then the client decides whether
to accept or decline the offer, etc.
Global graphs are a rather convenient formalism to express distributed
choices (as well as parallel computations) of work-flows.
As we mentioned before, an interesting feature of global graph is that they can easily show  branch/merge
points of distributed choices; for instance, in the global graph of Fig.~\ref{figure:global-graph} branching points merge in the loop-back node underneath the initial node.\\

Based on \defref{def:crn}, we can define two new binding mechanisms by exploiting the ``top-down'' (projection) and ``bottom-up'' (synthesis) nature offered by choreographies.
\begin{description}
\item[Top-Down]
  According to the first mechanism, provides-points are bound to
  requires-points when the projections of the global graph attached to
  the communication hyperarc are bisimilar to the corresponding
  communicating machine (exposed on the provides-points of services
  being evaluated for binding).
\item[Bottom-Up]
  The second mechanism is more flexible and it is based on a recent
  algorithm to synthesise choreographies out of communicating
  machines~\cite{lty15}.
  More precisely, one checks that the choreographies synthesised from
  the communicating machines, associated to the provides-points of
  services being evaluated for binding are isomorphic to the one
  labelling the communication hyperarc.
\end{description}

For example, the projections of the global graph of
Fig.~\ref{figure:global-graph} with respect to the components
$\processname{HotelsService}$ and
$\processname{PaymentProcessService}$ yields the communicating
machines in Figures~\ref{figure:communicating-machines}(b)
and~\ref{figure:communicating-machines}(c) respectively; so, when
adopting the first criterion, the binding is possible and it is
guaranteed to be well-behaved (e.g., there will be no deadlocks
or unspecified receptions~\cite{bz83}).
Likewise, when adopting the second criterion, the binding is possible
because the synthesis of the machines in
Fig.~\ref{figure:communicating-machines} yields the global graph of
Fig.~\ref{figure:global-graph}.

In this way, our approach combines choreography and orchestration by exploiting their complementary characteristics at two different levels.
On the one hand, services use global graphs to declare the behaviour expected from the composition of all the parties and use communicating machines to declare their exported behaviour.
On the other hand, the algorithms available on choreographies are used for checking the run-time conditions on the dynamic binding.

The resulting choreography-based semantics of binding guarantees
properties of the composition of services that are stronger than those
provided by the traditional binding mechanism of ARNs, and yielding a
more symmetric notion of interoperability between activities and
services.

\subsection{Comparison of the analysis and the binding mechanism}
Among the many advantages of developing software using formal tools,
is the possibility of providing analysis as a means to cope with
(critical) requirements. This approach generally involves the formal
description of the software artefact through some kind of contract
describing its behaviour. As we mentioned before, in SOC, services are
described by means of their contracts associated to their provides- and
requires-points, playing the role that in structured programming play
post- and pre-conditions of functions, respectively. From this point
of view, analysing a software artefact requires:
\begin{itemize}
\item the verification of the computational aspects of a service with
  respect to its contracts, yielding a \emph{coherence condition},
  whose checking takes place at design-time, and
\item the verification of the satisfaction of a property by an
  activity with respect to a given service repository, yielding a
  \emph{quality assessment} of the software artefact, whose checking
  takes place also at design-time.
\end{itemize}

On the other hand, service-oriented software artefacts require the
run-time checking associated to the \emph{binding mechanism}, in order
to decide whether a given service taken from the repository provides
the service required by an executing activity.

Table~\ref{tab:comparison} shows a comparison of the procedures that have to be implemented for checking the coherence condition of a service, the quality assessment of a service-oriented software artefact with respect to a particular repository, and for obtaining a binding mechanism for both of the approaches, the one based on ARNs, and the one based on CRNs.

\begin{table}[h!]
\begin{center}
\begin{tabular}{| c | p{3.5cm} | p{3cm} | p{4.5cm} |}
\hline
Formalisation & Coherence Condition & Quality assessment & Binding Mechanism\\
\hline
ARNs & $$\{\Delta_{\Lambda_p} \models^{\sf LTL} \Gamma_{\pi}\}_{\pi \in \gamma_p}$$ where $p \in P$, $\langle \gamma_p, \Lambda_p \rangle$ is a process, $\Delta_{\Lambda_p}$ the set of traces of the Muller automaton $\Lambda_p$ and $\Gamma_{\pi}$ is the LTL contract associated to port $\pi$. & $$\prod_{m \in P \cup C}\Lambda_m$$ & $$\Gamma_{\pi} \vdash^{\sf LTL} \Gamma_\rho$$ where $\pi$ is a provides point of a service, $\rho$ is a requires point of an activity, and $\Gamma_\pi$ and $\Gamma_\rho$ their LTL contract respectively.\\
\hline
\multirow{2}{*}{CRNs} & $$\{\Lambda|_{p_\pi} \approx \mathcal{A}_{\pi}\}_{\pi \in \gamma_p}$$ where $\Lambda|_{p_\pi}$ is the projection of Muller automaton $\Lambda$ over the alphabet of port $\pi$, $\mathcal{A}_{\pi}$ is the communication machine labelling port $\pi$ and $\approx$ denotes bisimilarity. & $$\prod_{m \in P}\Lambda_m$$ & {\bf Top-Down}: $$G|_\rho \approx \mathcal{A}_\pi$$ where $\pi$ is a provides point of a service, $\rho$ is a requires point of an activity, ${G_c}|_{p_\rho}$ is the projection of the global graph $G_c$ over the language of the port $\rho$, $\mathcal{A}_{\pi}$ is the communication machine labelling port $\pi$ and $\approx$ denotes bisimilarity.\\
& & & {\bf Bottom-Up}: $$S(\{\mathcal{A}_\pi\}_{\pi \in \Pi}) \equiv G_c$$ where $\Pi$ is the set of provides-points of the services to be bound, $G_c$ is the global graph associated to $c \in C$, $S(\bullet)$ is the algorithm for synthesising choreographies from communication machines~\cite{lty15} and $\equiv$ denotes isomorphism.\\
\hline
\end{tabular}
\end{center}
\caption{Comparison of the procedures for the approaches based in ARNs and CRNs}
\label{tab:comparison}
\end{table}


%% file: conc.tex
\section{Concluding Remarks}
\label{conc}

We propose the use of communicating relational networks as a formal model for service-oriented software design. CRNs are a variant of ARNs that harnesses the orchestration perspective underlying ARNs with a choreography viewpoint for characterising the behaviour of participants (services) over a communication channel. The condition for binding a provides-points of services to the requires-points of a communication channel of an activity relies on checking the compliance of the local perspective of the process, declared as communicating machines, with the global view implicit in the choreography associated to the communication channel. 
The binding mechanisms of ARNs (i.e., the inclusion of the set of
traces of the provides-point of the service bound in the set of traces
allowed by the requires-point of the activity) yields an asymmetric
acceptance condition.
Our approach provides a more symmetric mechanism based on
rely-guarantee types of contracts.

Our framework requires the definition of a criterion to establish the coherence among the Muller automaton $\Lambda$ of a process hyperedge and the communicating machines associated to its provides-points. This criterion, checked only at design time, is the bisimilarity of the communicating machine projected from $\Lambda$ and the ones associated to the provides-points. The reader familiar with Müller automata should note that defining such projection is not trivial when the automata are defined over a powerset of actions. The definition of the projection from Muller automata to communicating machine is conceptually straightforward (although technically not trivial) if the automata are defined over sets of actions (instead of powersets of them). Altough this is enough for the purposes of this paper, a better solution would be to extend communicating machines so to preserve the semantics of Muller automata even when they are defined on powersets of actions. This is however more challenging (as the reader familiar with Muller automata would recognise) and it is left as a future line of research.

We strived here for simplicity suggesting trivial acceptance
conditions.  For instance, in the ``bottom-up'' binding mechanism we
required that the exposed global graph coincides (up to isomorphism)
to the synthesised one.
In general, one could extend our work with milder conditions using
more sophisticated relations between choreographies.
For instance, one could require that the interactions of the
synthesised graph can be simulated by the ones of the declared global
graph.

We also envisage benefits that the orchestration model of ARNs could
bring into the choreography model we use (similarly to what suggested
in~\cite{bdft14}).
In particular, we argue that the 'incremental binding' naturally
featured in the ARN model could be integrated with the choreography
model of global graphs and communicating machines.
This would however require the modifications of algorithms based on
choreography to allow incremental synthesis of choreographies.
